# Wavelength calibration and spectral sensitivity correction of luminescence measurements for dosimetry applications: method comparison tested on the IR-RF of K-feldspar

Mariana Sontag-González[1], Dirk Mittelstraß[2], Sebastian Kreutzer[3,4], Markus Fuchs[1]

[1]Department of Geography, Justus Liebig University Giessen, 35390 Giessen, Germany

[2]Independent Researcher, Zschertnitzer Weg 16, 01217 Dresden, Germany

[3]Institute of Geography, Ruprecht-Karl University of Heidelberg, 69120 Heidelberg, Germany

[4]Archéosciences Bordeaux, UMR 6034, CNRS - Université Bordeaux Montaigne, 33607 Pessac Cedex, France

**Abstract**

Spectroscopic investigations provide important insights into the composition of luminescence emissions relevant to trapped-charge dating of sediments. Accurate wavelength calibration and a correction for the wavelength-dependent detection efficiency of the spectrometer system are crucial to ensure the correct spectrum interpretation and allow for its comparison with those obtained from other systems. However, to achieve an accurate detection efficiency correction, it is necessary to obtain the device-specific spectral response function (SRF). Here, we compare two SRF approximation methods by using either a calibration lamp of known irradiance or calculating the product of efficiency curves provided by the manufacturers of all known optical elements. We discuss the results using radiofluorescence (RF) measurements of two K-feldspar samples as an example. Feldspar infra-red (IR) RF spectra are known to be composed of several overlapping emissions, whose variation with sample mineralogy is still poorly understood and requires more extensive investigations. We find that both methods of sensitivity correction yield broadly similar results. However, the observed differences can alter a spectrum's interpretation. For example, we observe that after peak deconvolution the maximum signal wavelength of the IR-RF peak used for dating applications differs by ~3–13 nm between the two methods, depending on sample and diffraction grating. We recommend using calibration lamps to determine a device's SRF but highlight the need to consider issues such as higher-order signals in the choice of filters to establish the SRF's reliable wavelength range. Additionally, we find that a simple and inexpensive fluorescent white light yields an acceptable wavelength calibration comparable to that obtained from a specialized light source.

**Keywords**

*Emission spectroscopy, efficiency calibration, radiofluorescence, spectrum deconvolution*

**1 Background**

Luminescence signals, as obtained from typical dosimetric applications, are composed of individual emissions centred on different wavelengths. Typically, the bulk luminescence is quantified using a photomultiplier tube, assuming that the desired wavelength component dominates the signal within the detection wavelength range, which can be controlled through the selected optical filters. However, this approach misses the measured sample's composition of luminescence emissions. Information on the emission wavelengths can inform on the type of defects in the crystal lattice responsible for the luminescence and the characteristics of luminescence production (e.g., Krbetschek et al., 1997; Erfurt, 2003; Lomax et al., 2015; Friedrich et al., 2018; Schmidt and Woda, 2019; Niyonzima et al., 2020; Kumar et al., 2020; Riedesel et al., 2021). This information is particularly valuable in the characterisation of new dosimeters but can also be relevant if only one of two overlapping emissions needs to be quantified or if heterogeneous materials are investigated, such as those sometimes found in sediment dating applications.

Spectra of thermoluminescence (TL), optically stimulated luminescence (OSL), and radiofluorescence (RF) emissions can be obtained from a spectrometer either integrated within commercially available luminescence readers (e.g., Bos et al., 2002; Richter et al., 2013; Yoshizumi and Caldas, 2014) or using independent custom-built equipment (e.g., Martini et al., 2009; França et al., 2019). However, even when using integrated equipment, several application-specific issues need to be addressed by the end-users. A crucial issue is the wavelength-dependent difference in detection sensitivity of different emissions: the signal intensities of two neighbouring signal peaks are not directly comparable if their detection efficiency is not the same. The wavelength dependence of the detection efficiency is dictated by the transmission and reflectance of the optical elements (i.e., filters, optics, light guide, spectrograph) and the detector's quantum efficiency (i.e., CCD), collectively referred to as spectral functions. The product of all individual spectral functions yields a single build-up-specific spectral response function (SRF) which describes the overall system's sensitivity. The SRF affects the measured (uncorrected) spectrum as stated in Eq. (1), where the background spectrum can be caused by, e.g., intruding ambient light or dark signal produced by the camera.

$$measured\ spectrum = luminescence\ spectrum * SRF + background\ spectrum \qquad (1)$$

To get an approximation of the true luminescence (i.e., an efficiency calibration), we need to rearrange Eq. (1):

$$luminescence\ spectrum = \frac{measured\ spectrum - background\ spectrum}{SRF} \qquad (2)$$

One common approach to obtaining the SRF is to estimate it from the individual spectral functions supplied by the manufacturers of each optical element and the detector. We refer to this approach as the 'theoretical SRF', as, in theory, the product of the individual components' sensitivity yields the SRF of the entire system. However, this assumes that (i) the end-user is able to quantify every optical element in the system with a significant impact on the measurement, (ii) that the necessary datasheets are available and include the wavelength range of interest and (iii) that inter-lot sensitivity variations and degradation effects are insignificant for all optical elements. Alternatively, the SRF can be determined empirically using a reference light source with known radiance, in which case no further assumptions need to be made.

Here, we will use both approaches to obtain the SRF for a commercially available system for measuring RF spectra, integrated within a *lexsyg research* luminescence reader located at the Justus Liebig University of Giessen. However, the same workflows can be applied to other luminescence emissions and other luminescence readers, given that the spectrometer setup is similar. We will also discuss whether the simpler method of obtaining the SRF from manufacturers' datasheets is sufficient for calibrating other spectrometer setups. Light disturbances such as stray light from the spectrograph and diffraction signals of higher order are not considered in Eq. (1) and (2). We will discuss diffraction signals of higher order in sections 2 and 3.2, but not other possible light disturbances. It is the practitioner's responsibility to take these into account and ensure that they are negligible or corrected.

## 2 Equipment and analytical methods

All measurements were performed using a Freiberg Instruments *lexsyg research* reader (Richter et al., 2013) controlled by LexStudio2 (v1.7.4) software. For a scheme of the spectrometry build-up, see Fig. 1. The reader is equipped with an annular $^{90}$Sr/$^{90}$Y beta source (0.061 Gy s$^{-1}$, determined from the standard RCQ batch 126) and a spectrometry detection unit consisting of an Andor *Shamrock 163* Czerny-Turner type spectrograph coupled to an Andor *Newton DU920P* back-illuminated charge-coupled device (CCD) camera (1024x255 pixel). Luminescence light was collected by a fused silica lens with 1x magnification and transmitted to the spectrometer through a quartz fibre optic bundle of 2.4 mm diameter. This combination of lens and light guide restricts the measurable area of the sample to a spot of around 2.4 mm diameter in the middle of the sample. The light guide fibre bundle is rearranged to a rectangle at the output side and directly coupled to the spectrometer. We kept the adjustable entry slit of the spectrometer nearly closed during wavelength calibration to increase the wavelength resolution but opened it for the other measurements to increase signal intensity. The spectrograph has an aperture of f/3.6, which equals an opening angle of 15.8° and thus a numerical aperture (NA) of 0.138. Because the spectrograph's NA is lower than the NA of any other optical element, the light yield can be considered to be limited by the spectrograph.

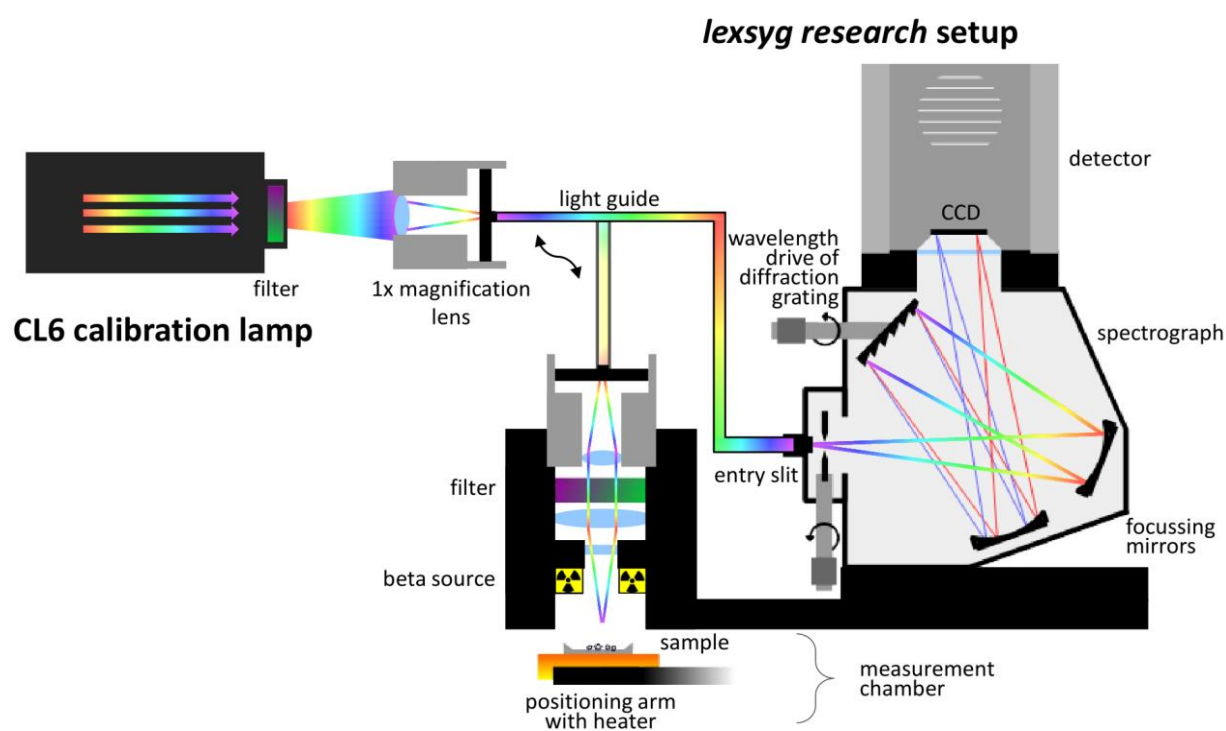

**Figure 1:** Diagram of the calibration lamp and the *lexsyg research* luminescence reader. The light guide can be directed towards the calibration lamp, which is positioned ~1–2 m away from the reader, or connected atop the RF position within the reader, as shown by the black arrow. Component sizes are not to scale.

The light diffraction in the spectrograph was achieved through a saw-toothed shaped mirror (blazed reflection grating, or, simply, a diffraction grating) manufactured by Richardson Gratings (Palmer, 2020). We tested two diffraction gratings: one with a groove density of 150 lines per mm (i.e., the density of 'teeth' in the saw-shape) and a blaze wavelength (i.e., the wavelength with maximum efficiency) of 800 nm (later referred to as 150/800 grating) and another grating with a groove density of 300 lines/mm and a blaze wavelength of 500 nm (300/500 grating). A higher groove density means light is diffracted to a larger angle, resulting in a smaller wavelength range but higher wavelength resolution. The angled mirror surfaces in a diffraction grating that allow for the spatial separation of light of different wavelengths also cause the diffraction of a single wavelength to multiple specific angles, leading to higher order signals in the measured spectrum. This means that diffracted light of, e.g., 300 nm will not only lead to a detected signal in the spectrum at 300 nm but also a second-order signal at 600 nm and a third-order signal at 900 nm, though the signal intensity decreases with increasing order. Importantly, the light maintains its original wavelength: the incoming light still has a wavelength of 300 nm, despite being detected at 600 or 900 nm as second- or third-order peaks. Thus, this signal can be suppressed if the detector has a low sensitivity at 300 nm or a suitable optical filter is used. Such higher-order signals must be taken into consideration when designing experiments or analysing spectral data. The SRF-based approaches presented here do not correct for higher-order signals, which are a possible source of measurement artefacts.

Our spectrometer build-up was optimized for detecting near-infrared radiofluorescence (IR-RF) of K-feldspar mineral grains for dating (e.g., Trautmann et al., 1998; Erfurt and Krbetschek, 2003a; Frouin et al., 2017). To detect IR-RF emissions centred around 865–880 nm (Trautmann et al., 1999; Kumar et al., 2018), a longpass (LP) filter with a cut-on wavelength of 500 nm (FELH0500, Thorlabs) was placed in front of the light guide, avoiding blue and UV emissions from the measurement chamber whose higher-order signals would appear in the wavelength range of interest.

To account for signal background due to ambient light and detector dark signals, a background spectrum was recorded for each experimental setting (i.e., for each used grating or setting). In the case of the calibration measurements, the lamps (calibration or fluorescent ceiling lamps) were not turned on, and in the case of the IR-RF measurement, the signal from an empty sample holder was acquired. The background spectra were subtracted from the measurements prior to any data processing.

For the empirical SRF determination (section 3.2), we used an equipment-tailored Bentham Instruments Ltd. Calibration irradiance standard (Type: CAL CL 6, SERIAL No: 16753/5, calibrated 2$^{nd}$ June 2014) as a light source. The irradiance output is specified by the manufacturer for wavelengths ranging from 300 nm to 2500 nm with <2 % uncertainty (at 2σ) of the absolute irradiance and up to ± 0.5 nm wavelength deviation. The bright illuminance of 6750 lx would overexpose the camera's CCD if directly connected, so the light source was positioned ~1–2 m away from the luminescence reader and pointed towards the spectrometer's light guide (Fig. 1). Depending on signal intensity, we adjusted the exposure time to avoid saturating the detector. Background measurements always used the same exposure time as the main signal measurements.

Data analysis was conducted in an **R** programming environment v4.0.2 (R Core Team, 2020). We used the lm() function for regression analyses and the approx() function for interpolating data points, both of the **R** base package 'stats'. When tabular information on optical elements' sensitivities was unavailable, datasheet images were digitized using the free software GraphGrabber v2.0.2 (Quintessa Ltd). The spectra of sediment samples shown in section 4 contained abrupt signal spikes caused by either cosmic rays or the beta source's bremsstrahlung. These signal spikes were removed by applying a running median of length 3 along the time axis (i.e., median value per channel of three subsequent spectra) iteratively 6 times. This procedure also reduces the general noise. To remove remaining outliers a final histogram-based algorithm (Pych, 2004) was used. All smoothing steps were carried out via the apply_CosmicRayRemoval() function of the 'Luminescence' (v0.9.19) package (Kreutzer et al., 2012, 2021).

**2.1 Wavelength calibration**

The output of the spectrometer is in signal intensity per each of the 1024 pixel columns of the CCD unit. A wavelength calibration consists of determining a polynomial function to transform the pixel position on the x-axes of the measured spectra into wavelength units (nm) and must be done for each spectrometer setup. Which wavelengths are diffracted onto which pixel column depends primarily on the chosen grating and the setting of the wavelength drive (a sine drive). The groove density of the former determines the range of spectral coverage, and the latter determines the centre wavelength of that range. In our setup, adjusting the wavelength drive rotates the grating within the stationary unit to shift the wavelength range; each 10 µm rotation shifts the central wavelength by 80 nm with the 150/800 grating or by 40 nm with the 300/500 grating. Here, the drive was rotated 165 µm and 260 µm for the 150/800 and 300/500 gratings, respectively.

For the wavelength calibration, we used the 'white-light' ceiling lamp in the laboratory, a common tubular fluorescent lamp (also known as 'neon tube'), as a wavelength standard. All commercially available fluorescent lamps use mercury and rare-earth phosphors containing the same elements yielding equivalent, reproducible peaks. The peak maximum locations correspond to ionisation light emission lines given in scientific databases (e.g., Kramida et al., 2021). The wavelength positions of the peak maxima are manufacturer-independent, but their intensities can vary depending on the proportion of containing phosphors (Elvidge et al., 2010).

It can be difficult to ascertain which emissions are dominant at certain wavelengths because some of the emission wavelengths are similar for different phosphors and the phosphor proportions are unknown, so we assume an uncertainty of ±5 nm for our calibration. Using the 150/800 grating, fourteen peaks could be identified as the known emissions of individual elements present in fluorescent lamps (Fig. 2a). Note that the last six peaks correspond to second- or third-order signals of the first five (e.g., the Hg peak expected at 435.8 nm appears repeated at 2x435.8=871.6 and 3x435.8=1307.4 nm). The actual calibration occurred through a third-degree polynomial regression line between the determined pixel positions and the known wavelengths of the fourteen identified peaks (Fig. 2b). The fitted parameters were inserted into the software *LexStudio2* (Clicking path: Advanced → Spectrometer → Calibration) to obtain wavelength-calibrated spectral measurements or used later as a data analysis step. Only ten peaks were identified using the 300/500 grating, which has a smaller wavelength range (Fig. 2c). Note that the last two peaks (871.6 nm and 976.8 nm) correspond to second-order signals, even though the first-order peaks (435.8 nm and 488.4 nm) are not measured.

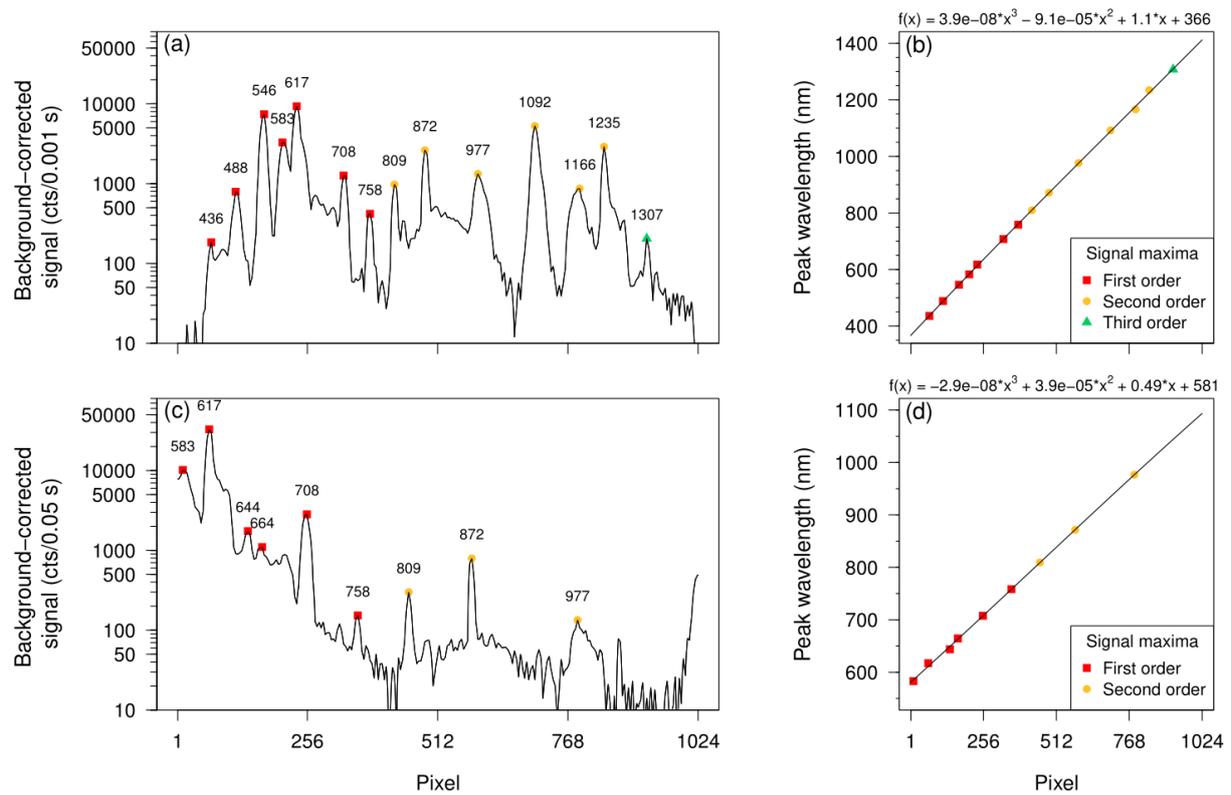

**Figure 2:** Wavelength calibration of the spectrometer with (a, b) a 150/800 or (c, d) a 300/500 grating. (a, c) Measured spectra of the fluorescent lamp labelled with wavelengths (in nm) of identified ionisation light emission lines (Kramida et al., 2021). (b, d) Fitting of pixel positions of signal maxima from the previous panel with known peak wavelengths using a third-degree polynomial (fitted parameters shown at the top).

We checked the accuracy of our fluorescent lamp calibration approach by repeating the calibration procedure with a QuantumDesign Hg(Ar) Pen-Ray line source manufactured especially for wavelength calibration. Such specialized light sources are the preferred method for spectroscopic studies but might not be available to all practitioners. We found that the differences between both calibrations in the wavelength range 500–1000 nm are at most 5 nm (Fig. 3, further details given in Appendix A). A third possible calibration method is to use the built-in 'solar light simulator' (SLS) LEDs in the *lexsyg research*, but we find that these lead to results which are not in accordance to the other two methods. The different maximum powers of the LEDs lead to large intensity differences between the LED peaks (see figure A2) which suggests that further data processing is needed. In addition, LED peak wavelengths are affected by temperature effects and other factors (e.g., Chhajed et al., 2005). They are, therefore, not suitable for reproducible and precise wavelength calibrations. However, the LEDs of the SLS can be used to generate a spectral fingerprint and can thus be an easy method to check which grating is built-in and if the spectrograph and filter settings are as expected.

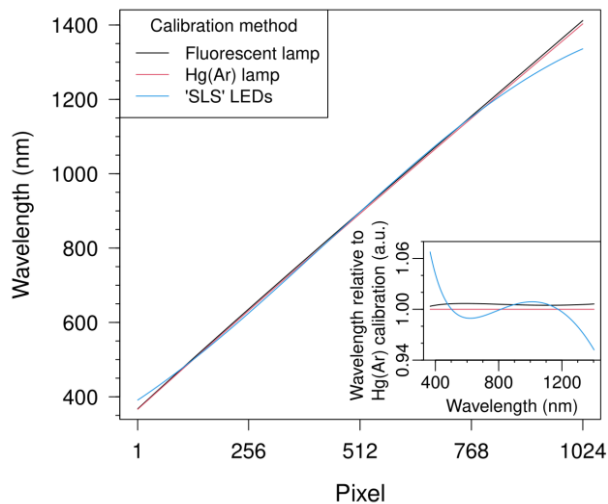

**Figure 3:** Comparison of three wavelength calibration methods. The lines correspond to the third-degree polynomials fitted to the peaks of a white-light fluorescent lamp, a Hg(Ar) wavelength calibration lamp and the *lexsyg* 'solar light simulator' LED array ('SLS'); these are the same regression lines as shown in Fig. 2b, Fig. A1b and Fig. A2b, respectively. The inset shows the wavelengths obtained from each method divided by those obtained from the Hg(Ar) calibration lamp method.

## 3 Determining spectral response functions

### 3.1 Calculating a spectral response function from datasheets

Our first approach to estimate the SRF was to use the manufacturers' datasheets for the six optical elements identified in the optical build-up when performing IR-RF measurements (see section 2). The theoretical SRF ($S_{\text{total datasheet}}$) obtained by combining the datasheet relative sensitivities of the individual elements ($S_i$) was calculated following Eq. (3):

$$S_{\text{total datasheet}} = S_{\text{LP filter}} * S_{\text{fibre optic bundle lens}} * S_{\text{fibre optic bundle}} * S_{\text{spectrometer optics}} * S_{\text{diffraction grating}} * S_{\text{camera}} \quad (3)$$

Individual curves and the resulting $S_{\text{total datasheet}}$ are shown in Fig. 4. We extrapolated the curves where the individual functions were unavailable for the entire range of interest (shown as dashed lines in Fig. 4). For the sensitivity of the 300/500 grating outside of the range 300–800 nm, we extrapolated following a linear fit of the nearest 50 nm. In the other cases, we assumed the value of the nearest point. These approximations should be interpreted as maximum values. The manufacturer provides the spectral sensitivity of the diffraction gratings for two directions of incoming polarized light (parallel and perpendicular). Since we assume that the incoming light is not polarized, we calculated the average of the two curves (Palmer, 2020).

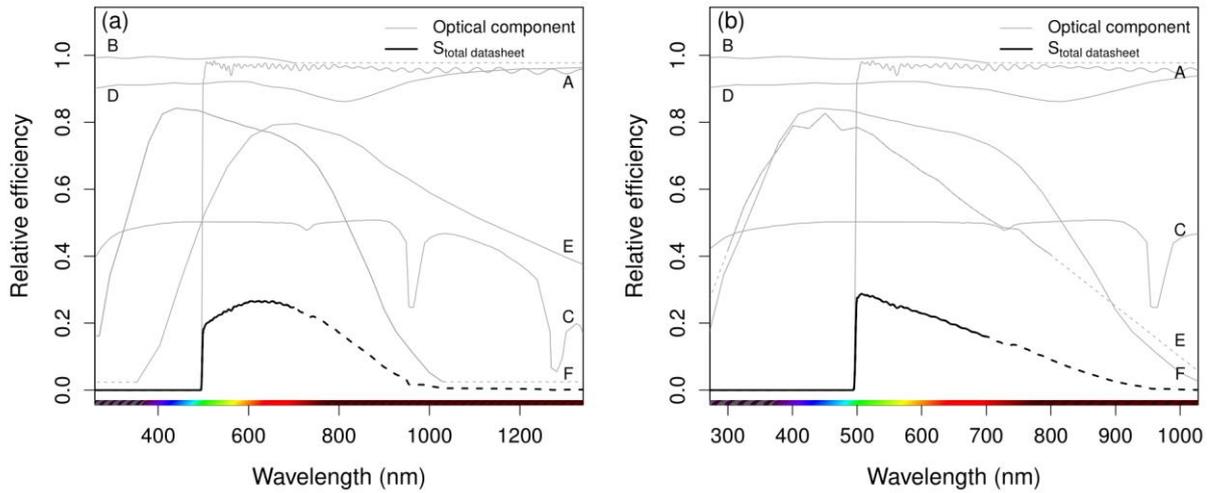

**Figure 4:** Spectral response (S) of the optical components in the spectrometric system using (a) a 150/800 grating and (b) a 300/500 grating (components are noted as A: longpass filter (500 nm), B: fused silica lens (1x magnification) at the entry of fibre optic bundle, C: quartz fibre optic bundle; D: coating of spectrometer optics, E: diffraction grating (average of s- and p-planes), F: CCD camera). The product of all individual spectral responses obtained from datasheets of the optical components yields $S_{\text{total datasheet}}$. If the individual functions were unavailable for the entire range of interest, the relationship was extrapolated horizontally (components E and F in plot (a) and component B in plots (a,b)) or following a linear fit of the nearest 50 nm (component E in plot (b)); shown as dashed lines.

### 3.2 Empirically determining a spectral response function

Our second approach to estimate the SRF was to use the CL6 calibration standard lamp described in section 2. The calibration lamp has a known light intensity per wavelength provided in the *Certificate of Calibration*. The light intensity was provided as energy flux (in units eV s$^{-1}$ m$^{-2}$ nm$^{-1}$), while the CCD camera output is the photon flux (in units cts s$^{-1}$ m$^{-2}$ nm$^{-1}$). Thus, we converted the reference spectrum by dividing the light intensity by the photon energy at each wavelength to obtain the expected measurement spectrum of the lamp (dashed lines in Fig. 5). We measured the calibration lamp output with the system described in section 2 and two different diffraction gratings: 150/800 and 300/500 (continuous lines in Fig. 5a and b, respectively). The measured spectra clearly differ from the lamp's reference emission spectrum, supporting the necessity of such a calibration.

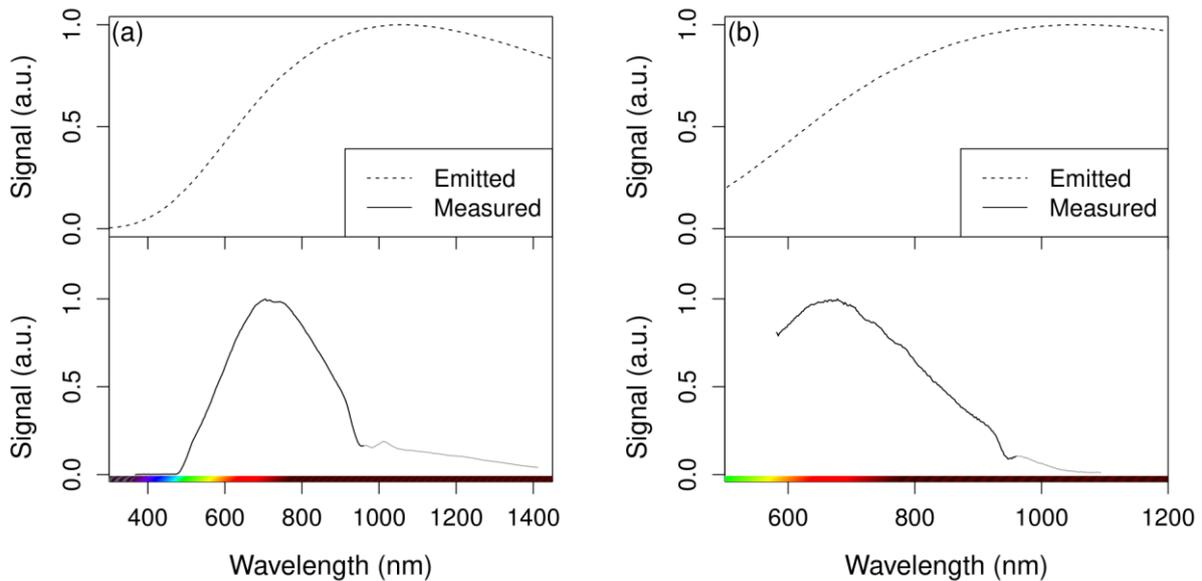

**Figure 5:** Calibration lamp spectrum according to the manufacturer (dashed lines) and empirically obtained spectra using (a) a 150/800 grating and (b) a 300/500 grating. The wavelength regions containing second-order signal (i.e., above 960 nm) are shown in grey colour.

The reference lamp has a broad emission wavelength range beginning at about 400 nm (see dashed line in Fig. 5a). Therefore, it is essential to consider the effect of higher-order signals in measurements. We used a 500 nm LP filter which according to the manufacturer transmits only wavelengths >508 nm with high efficiency and blocks wavelengths <492 nm. Wavelengths of 492–508 nm are partly transmitted. Higher-order signals appear at a multiple of the original wavelength, so the lowest transmitted wavelength of ~492 nm would appear again at ~984 nm (and at ~1476 nm, etc.). This is apparent in Fig. 5a, where we would expect negligible signal above 1000 nm due to the extremely low efficiency of the CCD camera (see

Fig. 4a, line F). Since higher-order signals are detected with the efficiency of the original wavelength, our measurements yield non-zero 'parasitic' signals even at wavelengths where the camera is blind, as exemplified in Fig. B1. This phenomenon sets an upper limit of our efficiency correction to twice the value of the lowest transmitted wavelength, which in our case would intuitively be 984 nm. However, the transmission boundary wavelengths are only valid for normal light incidence; with increasing incidence angle, the transmitted bandpass of an interference filter becomes broader (e.g., Lissberger and Wilcock, 1959). In our setup, we observed signal transmission beginning already at ~480 nm (see Fig. 6a and Fig. B1), which we attribute to imperfect collimation of the sample light (achieved by a lens placed atop the beta source; see Fig. 1). By consequence, our calibration has an upper limit of 960 nm. Depending on the wavelength range of interest, the upper limit could be changed by using a different filter, e.g., the 850 nm LP used in Fig. B1.

SRFs were then determined by dividing the measured spectrum for each grating by the lamp's reference photon emission spectrum. The resulting SRFs ($S_{\text{total CL6 lamp}}$) are shown in Fig. 6 together with the datasheet-based SRFs ($S_{\text{total datasheet}}$) obtained in section 3.1. Overall, we observe good agreement between the two curves obtained by the different methods. However, a few discrepancies are apparent: (i) the 500 nm LP filter does not have such a sharp beginning of transmission as expected from the datasheet information but a gradual change ranging 480–525 nm (Fig. 6a), presumably caused by the incidence angle of the sample light; (ii) a slightly higher relative efficiency of $S_{\text{total datasheet}}$ above 800 nm with the 300/500 grating (Fig. 6b), which can be explained by our very rough assumption of a linear decreasing efficiency of the grating above 800 nm (this wavelength range was not reported by the manufacturer, see line E in Fig. 4b); (iii) for both gratings, there is a discrepancy in the efficiency at ~950 nm between $S_{\text{total CL6 lamp}}$ and $S_{\text{total datasheet}}$. We conclude that this discrepancy is caused by an incorrect theoretical transmission function of the fibre bundle (line C in Fig. 4) as the wavelength range of the deviation matches that of the reported absorption band. Whereas the absorption band is also present in the CL6 measurements, it appears to be less pronounced than suggested by the datasheet. In addition, it should be noted that above ~900 nm small deviations between the theoretical and real spectral functions are intensified by the low overall efficiency.

We also recommend restricting the wavelength range of SRFs to regions with a minimum sensitivity (e.g., 500–1000 nm), because very low SRF values can amplify even low levels of signal noise.

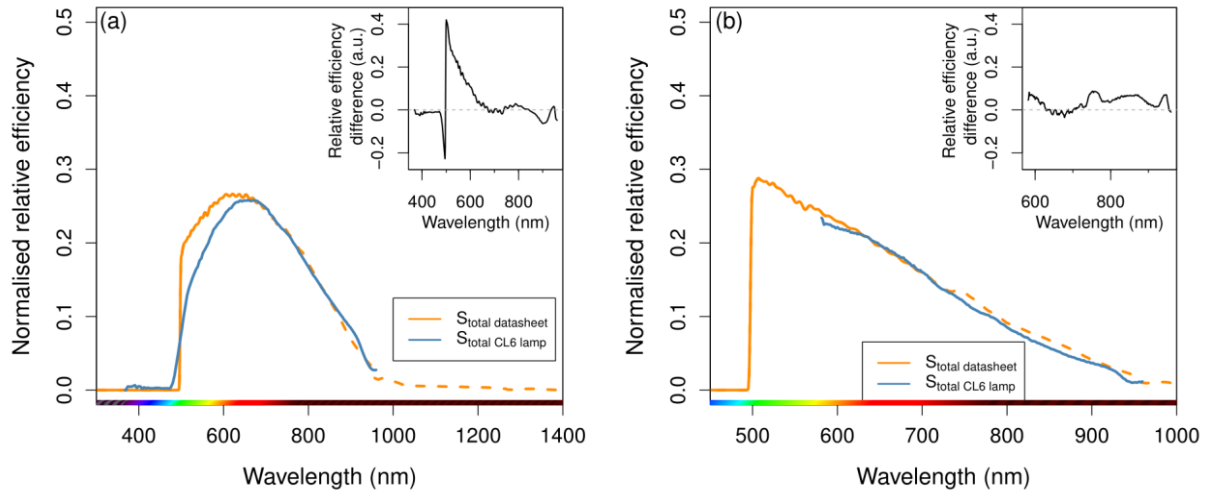

**Figure 6:** Comparison of methods to obtain SRF using (a) a 150/800 grating and (b) a 300/500 grating. $S_{\text{total CL6 lamp}}$ was obtained by dividing the measured spectrum by the calibration lamp's emitted spectrum (curves shown in Fig. 5). $S_{\text{total datasheet}}$ is reproduced from Fig. 4; dashed lines represent the wavelength range for which one or more spectral responses had to be extrapolated. $S_{\text{total CL6 lamp}}$ curves were normalized to the relative transmission value of $S_{\text{total datasheet}}$ at 700 nm. The insets show the difference between $S_{\text{total CL6 lamp}}$ and $S_{\text{total datasheet}}$.

## 4 Using the SRF for efficiency calibration

Once the SRF is estimated with either method (yielding $S_{\text{total datasheet}}$ or $S_{\text{total CL6 lamp}}$), it can be used to correct measured spectra by applying Eq. (2). We used the **R** function apply_EfficiencyCorrection() of the 'Luminescence' package to perform the correction, which has the advantage of interpolating mismatching x-axes and thus enabling the use of spectra with a different wavelength calibration or pixel binning than used for the efficiency measurement.

RF spectra obtained from two K-feldspar samples are shown as measured and after correction in Fig. 7 for both gratings. Samples Gi361, taken from a modern sand dune in Cuddalore, south-east India, and Gi326, taken from a Triassic sandstone near Bayreuth, Germany, were chosen for their bright IR-RF emission and known mineralogy (see Sontag-González and Fuchs (2022) for preparation details). The samples were measured as ~4 mm aliquots of grains of 90–200 µm in diameter mounted on stainless steel cups with silicone oil. Previously measured aliquots were fully bleached (25000 s with the in-built SLS), and after a 2 h pause, RF was measured at 70°C. Fig. 7 shows the spectra obtained from 19 s channels (~1.2 Gy) after ~3100 Gy regenerative dose for samples Gi361 and Gi326 using the 150/800 and 300/500 gratings. The SRFs were restricted from 525 nm to 960 nm. We set the lower limit of 525 nm to account for the transition range of the 500 nm LP filter transmission and the upper limit of 960 nm to cover only the first-order wavelength range transmitted through the filter.

The spectral sensitivity corrections led to similar results regardless of correction method, grating or sample. Compared to the uncorrected spectra, a small shift towards longer wavelengths could be observed for the IR-RF peaks, as well as an increase of the peaks' full width at half maximum (FWHM) using either grating, see Fig. 7. Additionally, for sample Gi326 a broad low-intensity peak in the red (~700 nm) almost disappears after the correction, whereas it is more pronounced in sample Gi361 and remains apparent after correction. Such a peak has been reported previously (e.g., at ~1.77 eV; Erfurt and Krbetschek, 2003b), and its possible contribution to the IR-RF signal measured with a photomultiplier tube might be a source of error for equivalent dose estimation due to the instability of this emission (Trautmann et al., 1998; Murari et al., 2021).

However, the calibration lamp method (using $S_{\text{total CL6 lamp}}$; blue spectra in Fig. 7) led to significant differences in specific peak attributes when compared to the datasheet method (using $S_{\text{total datasheet}}$; orange spectra in Fig. 7): (i) no unexpected signal increase at ~950 nm, (ii) a less steep signal decline in the range 900–950 nm with the 300/500 grating, and (iii) a less pronounced shift of the IR-RF peak maximum wavelength. These differences arise from those observed in the SRFs of both methods (see section 3.2).

In consequence, the maximum peak wavelengths differed for both correction methods. To gather statistical data, we analysed 1500 consecutive spectra of both samples, for each of the two gratings, obtained from automated measurement sequences. For both samples, the median wavelengths of the IR-RF peak maxima differ by ~13 nm and ~25 nm for the 150/800 and 300/500 gratings, respectively. Since low-temperature spectroscopic studies suggest that this IR-RF peak is composed of two emissions (Kumar et al., 2018; Riedesel et al., 2021), spectral deconvolution is useful to discuss differences in peak positions.

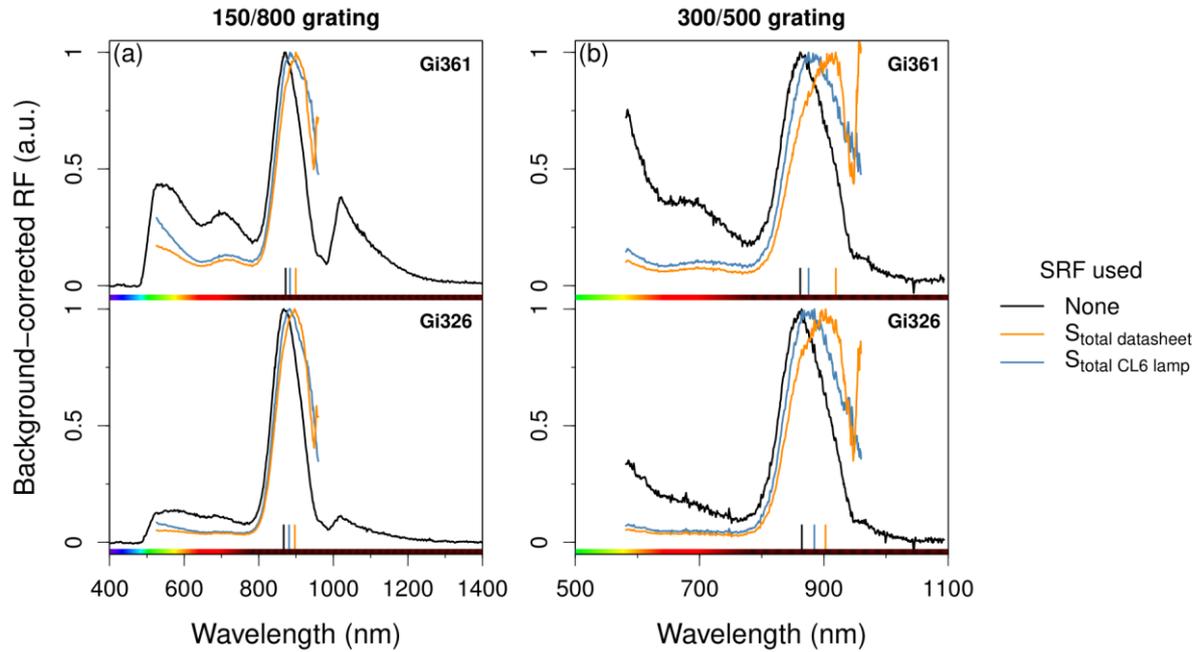

**Figure 7:** Radiofluorescence (RF) spectra of samples Gi361 and Gi326 (top and bottom plots, respectively) shown before and after efficiency calibration with two spectral response functions (SRF) using a 500 nm LP filter and (a) a 150/800 and (b) a 300/500 grating. The spectra were obtained after administering doses of ~3100 Gy. Spectra are normalised to the highest signal intensity of the IR-RF peak. Wavelengths of peak maxima are marked at the bottom of the plots.

**4.1 IR-RF peak deconvolution**

We also analysed the effect of the two correction methods on the resulting IR-RF peak positions after deconvolution of the 1500 consecutive spectra of both samples, for each of the two gratings. We only fitted the range 1.31–1.65 eV (750–950 nm) to focus on the emissions in the near infra-red and avoid the presumed artefact of signal increase beyond 950 nm when using $S_{total\ datasheet}$, which is presumably caused by the quartz fibre optic bundle efficiency curve. After the data pre-processing described in section 2, we converted the spectra's wavelength axes to energy (including a Jacobian correction) using the **R** function convert_Wavelength2Energy() of the 'Luminescence' package and then subtracted the RF value of the highest-energy channel (1.65 eV) of each spectrum as background. Finally, we fitted each spectrum with a sum of two Gaussian functions using the function fit_EmissionSpectra(). As shown in an example spectrum of each grating for each sample in Fig. 8, we achieve a good fit with the measured data using either correction method. However, the peak maximum wavelengths of the higher wavelength (lower energy) IR peak should be considered rough estimates, as only a portion of the peaks was fitted.

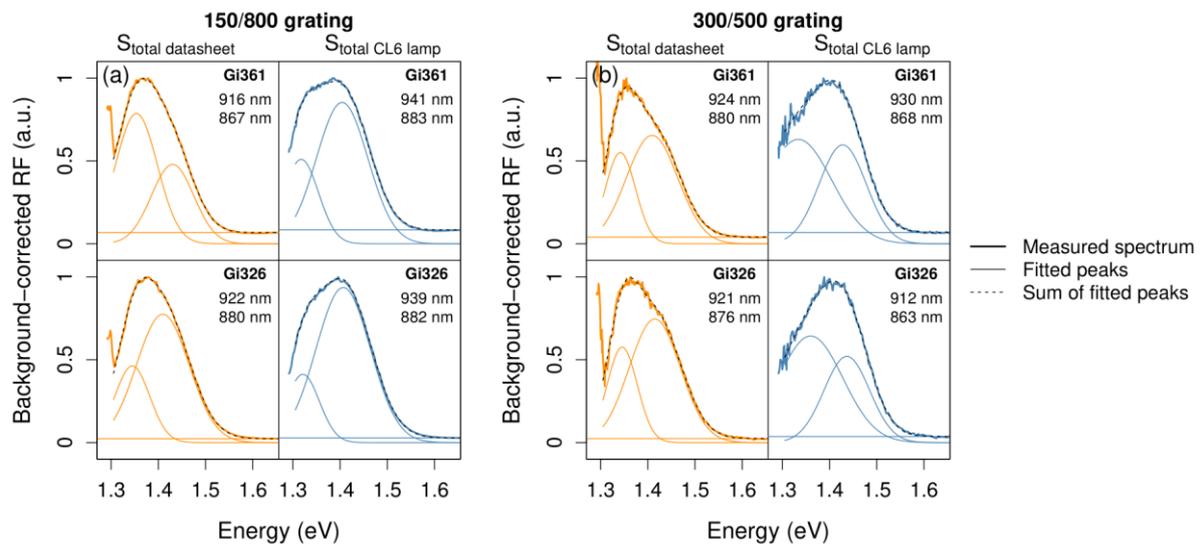

**Figure 8:** Deconvolution of example radiofluorescence (RF) emission spectra of samples Gi361 and Gi326 (top and bottom plots, respectively) after efficiency calibration measured with (a) a 150/800 and (b) a 300/500 grating. For each grating, the spectra were corrected using two spectral response functions (left- and right-hand plots). After background subtraction (horizontal lines), spectra were fitted with a sum of two Gaussian functions (wavelengths of peak maxima given in the individual legends) and normalized to the highest signal intensity of the IR-RF peak.

The distributions of the fitted peak energies for the lower wavelength (higher energy) IR peak, which is considered to be the main emission used for IR-RF dating (e.g., Murari et al., 2021), are shown in Fig. 9a–d for both samples. Though the distributions all partly overlap, peak maximum energies obtained with the 150/800 grating and the $S_{\text{total CL6}}$ correction (median values of 1.405 eV (882 nm) and 1.406 eV (882 nm) for Gi361 and Gi326, respectively) are slightly lower than those obtained with the 300/500 grating (median values of 1.427 eV (869 nm) and 1.433 eV (865 nm) for Gi361 and Gi326, respectively). The data from the $S_{\text{total datasheet}}$ correction showed no clear pattern, but median values range between 1.409 eV (880 nm) and 1.424 eV (871 nm).

The distributions of fitted peak maximum energies for the higher wavelength (lower energy) IR-RF peak are shown in Fig. 10. For this peak, the distributions differ markedly between the two sensitivity corrections and between the two gratings. For both samples measured with the 150/800 grating, there is no significant overlap between the distributions resulting from the two corrections. The median peak energy resulting from the $S_{\text{total datasheet}}$ correction is ~1.350 eV (~920 nm) and that from the $S_{\text{total CL6}}$ correction is ~ 1.320 eV (~940 nm) for both samples. Interestingly, the shape of the distribution of the $S_{\text{total CL6}}$ dataset for the 300/500 grating is much broader for both samples (blue histograms in Fig. 10c–d) than all the other distributions. No significant differences were expected between the two gratings, other than those caused by the lower wavelength resolution of the 150/800 grating. The observed high variability in the fitting of the $S_{\text{total CL6}}$ dataset for the 300/500 grating is presumably due to uncertainty in the fitting procedure caused by the low sensitivity in this wavelength range and the lack of information on the low-energy side of the peak.

Table 1 contains the interquartile ranges (the central 50% of a distribution) of peak maximum wavelengths for both IR peaks of each dataset. These ranges are broadly consistent with previous observations by Erfurt and Krbetschek (2003b) of 865 nm and 910 nm, by Kumar et al. (2018) of 880 nm and 955 nm and by Riedesel et al. (2021) of 874–885 nm and 917–953 nm (five samples of single crystal low temperature K-feldspar) and Kreutzer et al. (2022) of 868-871 nm (interquartile range for one sample, $n_{\text{fit}}$ = 528, 300/500 grating). When using the recommended empirical SRF ($S_{total\ CL6}$) and the 150/800 grating, which led to narrower distributions, our median peak wavelengths for the lower-wavelength IR emission of 882 nm (both samples), and for the higher-wavelength IR emission of 941 nm (Gi361) and 939 nm (Gi326) were more similar to those of Kumar et al. (2018) and Riedesel et al. (2021). Additionally, it should be noted that the spectra shown in Erfurt and Krbetschek (2003b) were not efficiency-corrected, so, based on our data, the median peak wavelengths can be expected to be 11–24 nm underestimated. Thus, we recommend that the values obtained by Kumar et al. (2018) be used to refer to the IR-RF emissions until more detailed studies involving more samples are undertaken. Based on the results of our two samples, when using a photomultiplier tube for IR-RF dating, a bandpass filter of longer wavelength (e.g., centred around 880 nm) would be more appropriate to capture the main IR-RF emission and avoid contaminating peaks below 800 nm than the 850/40 nm filter typically used in *lexsyg research* devices.

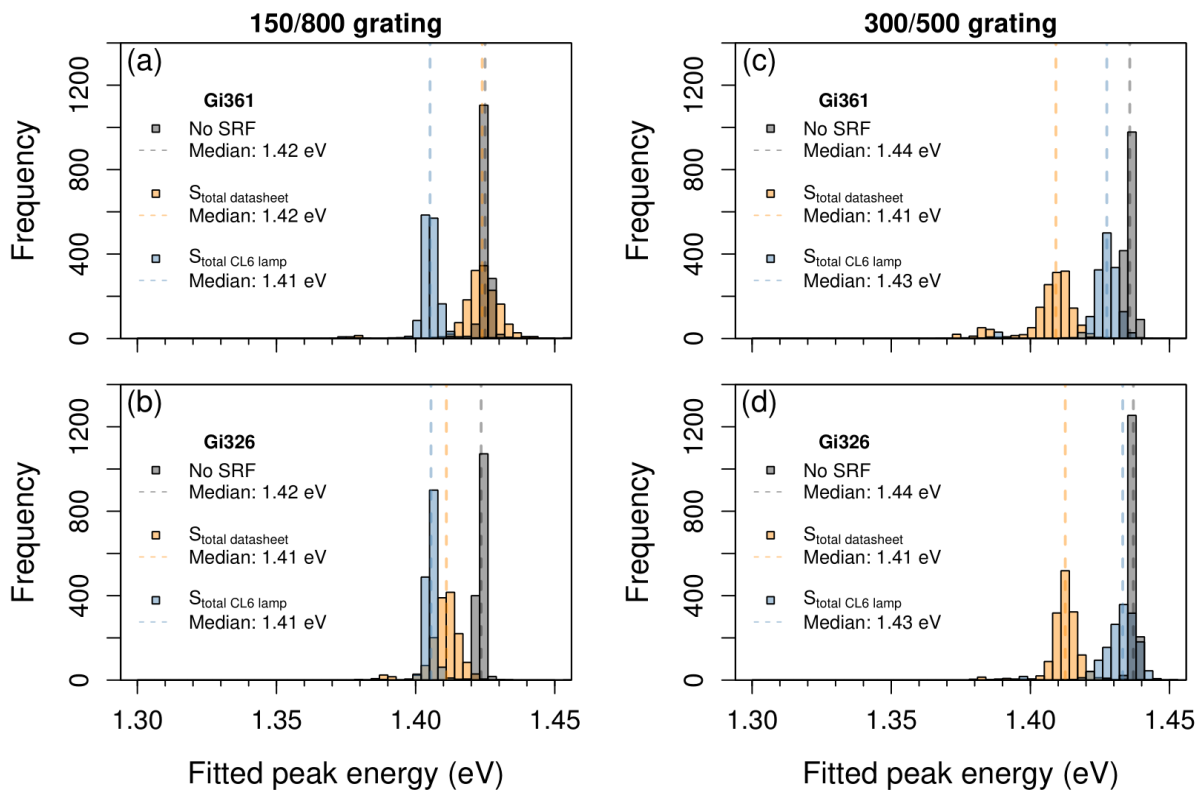

**Figure 9:** Histograms of fitted IR-RF peak energies for the lower-wavelength IR peak resulting from deconvolution of spectra of samples Gi361 and Gi326 (top and bottom plots, respectively) after either no efficiency calibration or with one of two methods. Spectra were obtained using a 500 nm LP filter and (a–b) a 150/800 and (c–d) a 300/500 grating. After background subtraction, spectra were fitted with a sum of two Gaussian functions. Up to 6 energies per dataset fall outside the plotted limits; these are considered outliers.

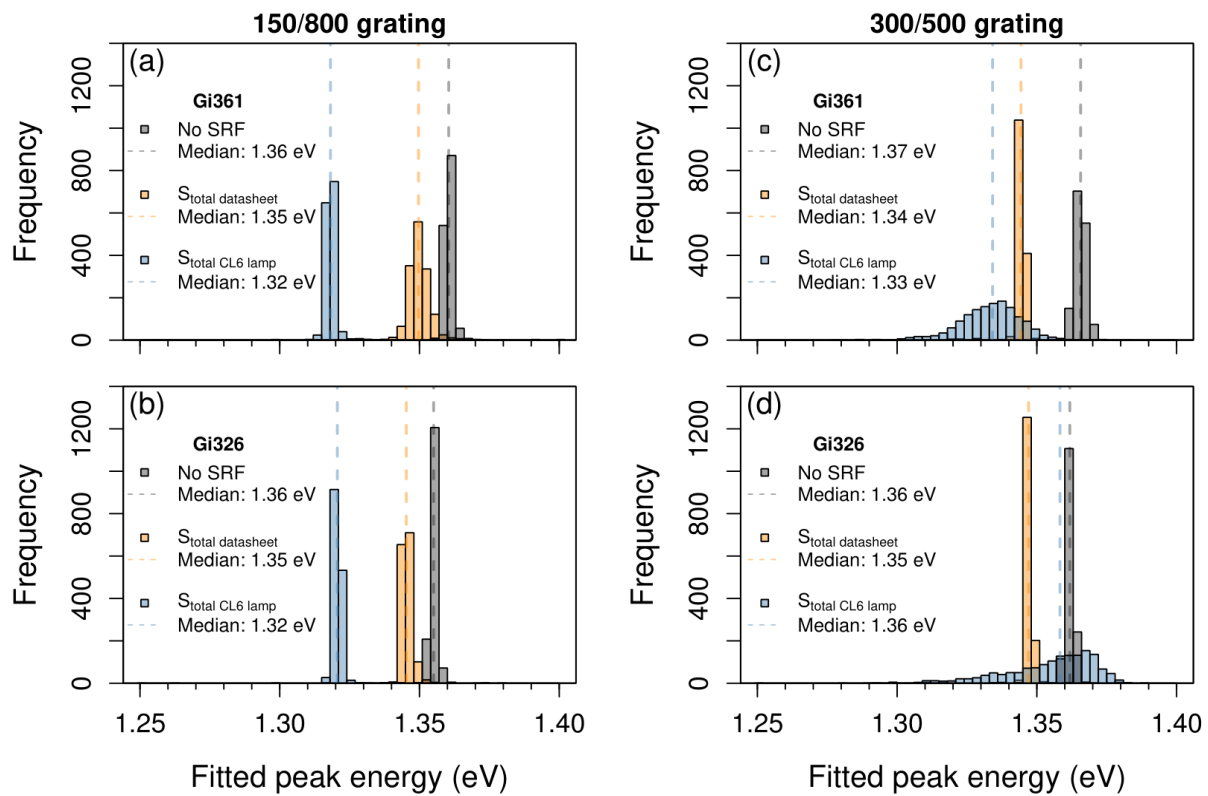

**Figure 10:** Histograms of fitted IR-RF peak energies for the higher-wavelength IR peak resulting from deconvolution of spectra of samples Gi361 and Gi326 (top and bottom plots, respectively) after either no efficiency calibration or with one of two methods. Spectra were obtained using a 500 nm LP filter and (a–b) a 150/800 and (c–d) a 300/500 grating. After background subtraction, spectra were fitted with a sum of two Gaussian functions. Up to 6 energies per dataset fall outside the plotted limits; these are considered outliers.

**Table 1**: Interquartile ranges of IR-RF peak wavelengths after energy spectra deconvolution with two Gaussian functions between 1.31 eV and 1.65 eV (950 nm and 750 nm). Each dataset consisted of 1500 spectra. IR peaks labelled as 1 and 2 refer to the lower- and higher-wavelength peaks, respectively.

| IR peak # | Sample | Interquartile range of wavelengths from fitted peaks (nm) | | | | | |
|---|---|---|---|---|---|---|---|
| | | 150/800 nm grating | | | 300/500 nm grating | | |
| | | No SRF | $S_{total}$ datasheet | $S_{total\ CL6}$ | No SRF | $S_{total}$ datasheet | $S_{total\ CL6}$ |
| 1 | Gi361 | 870–871 | 868–873 | 881–883 | 863–864 | 878–883 | 867–870 |
| | Gi326 | 871–871 | 877–880 | 882–883 | 862–863 | 876–879 | 863–867 |
| 2 | Gi361 | 911–912 | 917–920 | 940–941 | 907–909 | 923–924 | 925–935 |
| | Gi326 | 915–915 | 921–922 | 938–939 | 910–911 | 921–922 | 907–922 |

## 5 Conclusions and recommendations

We compared three methods for the spectrometer wavelength calibration (Fig. 3). Specialized light sources, such as the Hg(Ar) Pen-Ray line source, are considered standard practice due to the well-defined luminescent components (mainly mercury) and the resulting narrow and intense emissions. However, we show a similar accuracy for our equipment with the fluorescent ceiling lamp present in most laboratories, which provides a cheaper and safer (no strong UV emissions) method for wavelength calibration. LEDs do not yield an accurate calibration due to their broad emission peaks with unstable peak maximum wavelengths. However, SLS spectra can be used as a rough indication of the grating setup and the pre-set wavelength calibration.

As expected, there are stark differences between the corrected and not corrected RF spectra of two K-feldspar samples (Fig. 7), e.g., the IR-RF peak is narrower and its maximum wavelength is underestimated if the measurement is not corrected for the spectral response of the optical build-up and the detector. Both presented correction methods led to slightly different SRFs, as discussed in section 3.2 (see Fig. 6). Due to these differences, the empirical determination of the SRF for a specific spectrometer setup is recommended instead of a calculation from available datasheets despite the more substantial analytical procedure required, but the use of the latter can be expected to give an acceptable approximation.

Since our instrumental setup was optimized to characterize IR-RF emissions, equivalent calibration measurements would need to be conducted to determine $S_{\text{total CL6 lamp}}$ for studies of, e.g., K-feldspar IRSL (main emission at 410 nm, blue) or quartz OSL (main emission at 365 nm, UV). Note that correction with the SRF merely normalizes the signal intensities at different wavelengths, making peak heights comparable. Additionally, signals of higher order (see Fig. B1) also need to be considered before conclusions can be drawn from the sensitivity-corrected spectra. We solved this issue through a 500 nm longpass filter to remove UV and blue emissions from the luminescence light and restricted the analysed wavelength range to twice the effective cut-off wavelength (i.e., 960 nm). Future work should increase the edge of the cut-off filter (e.g., 600 nm longpass) to extend the analysed wavelength range up to the limit imposed by the CCD camera sensitivity (~1000 nm).

Lastly, though the choice of calibration method undoubtedly affects the wavelength of a reported emission peak (as shown in section 4.1 for IR-RF peaks), we sometimes observed larger variations of deconvoluted peak maximum wavelengths between different consecutive channels than between different gratings or calibration methods (see Figs. 9 and 10). We, therefore, recommend that future RF emission studies analyse several spectra of the same sample (e.g., >30) and report descriptive statistics of the emission wavelengths (e.g., median and interquartile range). Moreover, such variability of the fitting parameters, which we attribute to fitting uncertainty, should be considered when spectrum deconvolution is used for dating. One possibility of reducing the fitting uncertainty is to fix the peak position and width for all spectra in a dataset, allowing only the amplitude (i.e., the parameter of interest for dating) to vary.

**Appendix A - Wavelength calibration**

We conducted two additional wavelength calibrations with the 150/800 grating using alternative methods to check the accuracy of the fluorescent lamp calibration suggested in section 2.1 of the main text. We used a Hg(Ar) Pen-Ray line source (LOT-Oriel, now Quantum Design GmbH), which produces narrow spectral lines primarily from the excitation of vaporized mercury. According to the manufacturer, expected emission peaks are at (253.7, 302.2, 312.6, 334.0, 365.0, 404.7, 435.8, 546.1, 578.0) nm. We positioned the lamp around 5 cm away from the spectrometer light guide opening and set the exposure time to 0.001 s/channel. The entry slit was kept almost closed to increase the wavelength resolution and prevent overexposure of the detector. The original and higher-order signals are visible as sharp peaks in Fig. A1a. We assigned 18 peaks to the lamp's known-wavelength emissions and used them to fit a calibration line (Fig. A1b).

We also tested using the built-in 'solar light simulator' (SLS) LEDs in the *lexsyg research* for the wavelength calibration. The spectrometer light guide (with the 1x magnification adapter) was connected to the 'special TL' position in close vicinity of the SLS unit position, and the LEDs were manually activated through LexStudio2 (clicking path in v1.7.4: Advanced → Debug → chamber unit → solar light simulator). The six colours (IR, red, amber, green, blue and UV, respectively centred at 850, 625, 590, 523, 462 and 365 nm) were independently measured by parameterizing the chosen colour with a control value of 35000 and the remainder with zero. The LEDs were switched on for 20 s, during which time a spectrum was recorded (exposure time of 0.001 s/channel). The resulting six spectra are shown in Fig. A2a. First-order and second-order signals are visible for all LEDs except for the IR, where only the first-order signal is measured, and the UV, where only the second-order signal is measured (the signal intensity of the UV third-order signal was too low). The ten resulting peaks were used with the corresponding wavelengths to fit a calibration line (Fig. A2b).

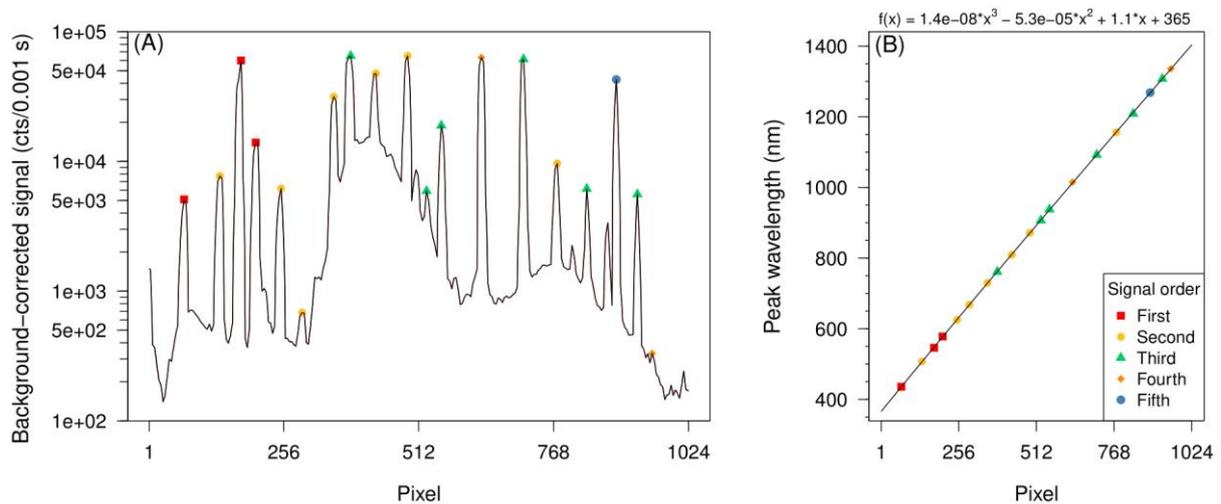

**Figure A1:** Wavelength calibration of the spectrometer (150/800 grating) using a Hg(Ar) Pen-Ray line source. (a) The measured spectrum with identified known peaks. (b) Fitting pixels of signal maxima from the previous panel with known peak wavelengths using a third-degree polynomial (fitted parameters shown at the top).

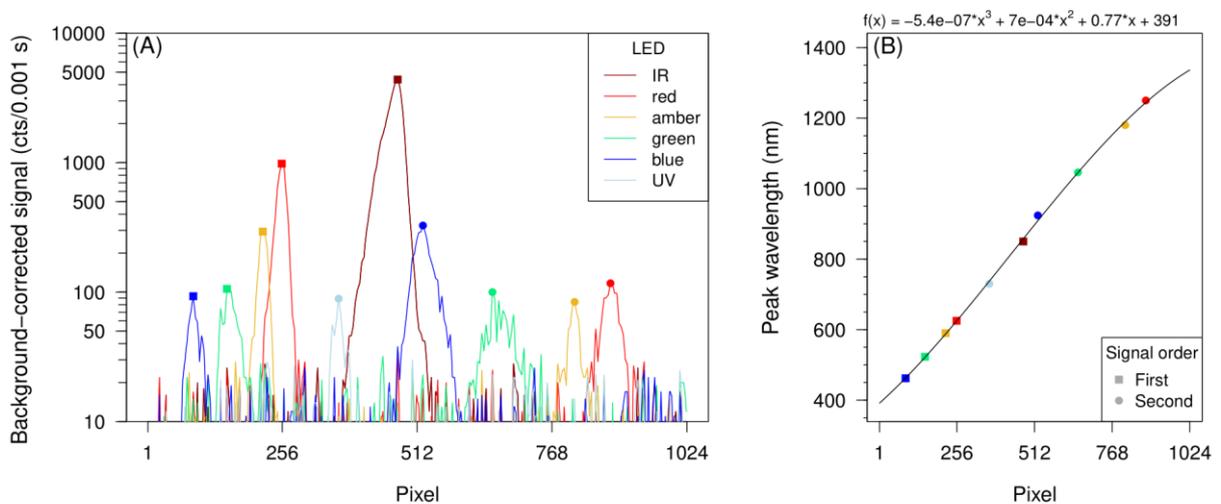

**Figure A2:** Wavelength calibration of the spectrometer (150/800 grating) using the built-in 'solar light simulator' LEDs in a *lexsyg research*. (a) Overlain measured spectra of LEDs (each colour measured

individually). (b) Fitting pixels of signal maxima from the previous panel with known peak wavelengths using a third-degree polynomial (fitted parameters shown at the top).

## Appendix B - Filter comparison

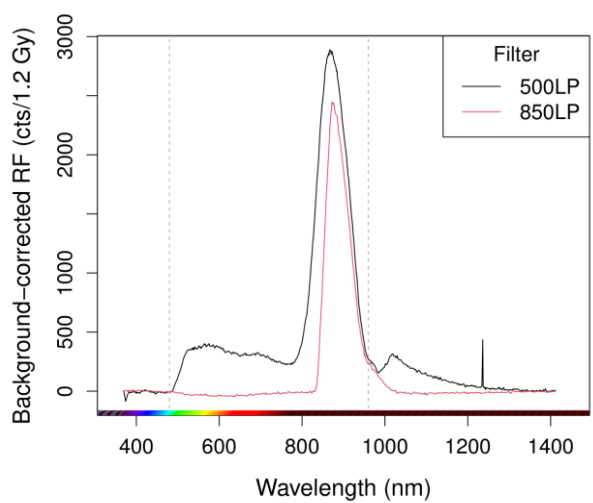

**Figure B1:** RF spectrum of sample Gi326 using a 150/800 grating and either a 500 nm or an 850 nm long-pass filter (500LP and 850LP) without a detection efficiency correction. The grey dashed vertical lines indicate the onset of transmission of the first and second-order signals using the 500LP (at 480 nm and 960 nm, respectively). The signal difference between the two curves at wavelengths >960 nm corresponds to a second-order signal. The spectra were obtained from the same aliquot after 2950 Gy regenerative dose (same aliquot as used in section 4).

## Data statement

The data and **R** code used for this work are available at https://doi.org/10.5281/zenodo.6565760 (Sontag-González et al., 2022).

## Acknowledgements

We are thankful to Andreas Richter and Freiberg Instruments for providing the irradiance reference lamp and the necessary technical information about the *lexsyg research* system. MSG and MF are supported by the German Research Foundation (DFG FU 417/36-1). The preliminary work by DM and SK was supported by the Federal Ministry for Economic Affairs and Climate Action (BMWK) based on a decision by the German Bundestag through a ZIM-KOOP Cooperation grant to the Justus Liebig University Giessen and Freiberg Instruments (KF3073002 DF3 & KF3072502 DF3). The LaScArBx supported the initial work of SK before 2020. LaScArBx is a research programme supported by the ANR (ANR-10-LABX-52). In the submission phase, SK received funding from the European Union's Horizon 2020 research and innovation programme under the Marie Skłodowska-Curie grant agreement No. 844457 (CREDit). We thank Johanna Lomax for revising the manuscript, Sumiko Tsukamoto for access to sample Gi361 and the three anonymous referees for their critical reviews.